\documentstyle[12pt,epsf]{article}
\newcommand{\includegraphics}[1]{\leavevmode\epsffile{#1}}

\begin{document}
\title{Intermittency and the Slow Approach\\ to Kolmogorov
Scaling}
\author{\normalsize B. Holdom\thanks{holdom@utcc.utoronto.ca}\\
\small {\em Department of Physics,}
\small {\em University of Toronto}\\
\small {\em Toronto, Ontario,}
\small M5S1A7, CANADA}
\date{}
\maketitle
\begin{abstract}
From a simple path integral involving a variable volatility in the
velocity differences, we obtain velocity probability density functions
with exponential tails, resembling those observed in fully
developed turbulence. The model yields realistic scaling exponents and
structure functions satisfying extended self-similarity. But there is an
additional small scale dependence for quantities in the inertial range,
which is linked to a slow approach to Kolmogorov (1941) scaling
occurring in the large distance limit.
\end{abstract}
\vspace{4ex}

The universal features displayed by fully developed hydrodynamic
turbulence are still not fully understood. Kolmogorov (1941)
\cite{ca,cb} showed how a set of statistical quantities known as
structure functions are expected depend on the length scale \(r\) as
power laws with predicted exponents. Experimental
measurements have indicated that while these predictions are close
to the truth, the predicted exponents are not exactly realized. In the
face of this experimental input, much effort has been devoted
towards understanding the origin of these anomalies in the scaling
exponents, while retaining the notion that current
experiments are observing an ``inertial range'' where strict power
law scaling holds.

In this paper we will investigate what appears to be a loop-hole in
this reasoning. In spite of impressive
advances made in the experimental studies,
the fact remains that the scaling exponents have been
deduced by looking at scaling regions extending over little more
than one decade in \(r\) \cite{cg,cf,ci,cc}. This leaves open the
possibility that a small but significant departure from strict power law
scaling is still consistent with the data. We will argue that this
possible departure is sufficient to allow the observed anomalies in
the exponents to be nothing more than a transient effect related to
intermittency, and that the true power law scaling only occurs on
larger scales. This very large distance scaling could take the form
proposed by Kolmogorov.

We shall present a model which predicts deviations from power
law scaling, and which shows that these deviations may be small
enough to have escaped detection thus far. The model is based on a
simple physical picture for the effects of intermittency, which are
effects due to the coming and going of coherent structures in the
velocity field. We will model the effects that these
structures (eddies) have directly on the probability distribution
function (PDF) for velocity differences. We are claiming simplicity
in our approach, but not uniqueness. Thus the applicability to
3-dimensional hydrodynamical turbulence in particular could be
considered speculative, since the model makes no mathematical
contact with the Navier-Stokes equations. On the other hand, when
the consequences of the model are explicitly worked out they are
found to coincide rather well with the data.

Only the symmetric part of the
observed PDF will be modeled in this paper. This is denoted by
\(P_r(\delta v_r)\) where \(\delta v_r\) is the difference in some
velocity component at two points separated by a distance \(r\). The
model will yield an explicit expression for the PDF which 1) for
small \(r\) has the typical sharp peak and broad tails characteristic of
intermittent behavior, 2) at any finite \(r\) displays
exponential tails at large enough \(\delta v_r\), and 3) tends toward a
Gaussian form in the large \(r\) limit. But the model shows that
these realistic features may in fact be implying that the values of the
scaling exponents are slightly scale dependent, as they evolve from
their observed ``anomalous'' values for the values of \(r\) where
they are presently measured, to the pure K41 values in the large \(r\)
limit.

We consider a discrete set of points on the line connecting the two
points separated by distance \(r\). We label the points by their
distance from one end, \((0,r_1,r_2,...,r_N,r_{N+1})\)  where
\(r_{N+1}=r\). We start by assuming that
\(P_r(\delta v_r)\) can be approximated by
\begin{equation} P_N( v_{r}-v_{0})=
\left[{\prod\limits_{l=1}^{N} \int_{-\infty }^{\infty
}dv_{r_l}}\right]{\cal P}( v_{r}-
 v_{r_{N}}){\cal P}( v_{r_{N}}- v_{r_{N-1}})\cdots {\cal P}(
v_{r_{1}}- v_{0}),
\label{bb}\end{equation} with \(\int_{-\infty }^{\infty }{\cal
P}(y)dy=1\). We assume that the set of points \(r_{j}\) is
characterized by a scale \(\rho\) such that \(({r}_{j}/\rho)^{a} = j\),
where \(j=1,2,...,N+1\) and \(a\) is a positive constant to be
determined below. The integration over all velocities at the
intermediate points may be thought of as a sum over all paths in the
space of velocities which start at \(v_{0}\) and end at
\(v_{r}\). We could thus refer to (\ref{bb}) as a path integral.  But
note that we refrain from taking the continuum limit,
\(N\rightarrow\infty \) with \(r\) fixed, since the scale
\(\rho\) has a physical significance.
 
The discretization of the range from 0 to \(r\) into subregions
\((r_{j},r_{j+1})\) is associated with our modeling of intermittent
behavior. Roughly speaking, we are suggesting that the coming and
going of eddies, of sizes
comparable to subregion size
\(r_{j+1}-r_{j}\), produce a {\em variable volatility} in the typical
velocity differences \(v_{r_{j+1}}-v_{r_{j}}\). We make this
statement precise by writing
\({\cal P}( v_{r_{j}}-v_{r_{j-1}})\) as a superposition of
Gaussians, where we integrate over all values of a volatility
parameter \(s_j\) which is itself weighted according to a Gaussian.
To simplify notation we denote
\( v_{r_{j}}=x_{j}\).
\begin{eqnarray} {\cal P}(x_{j}-
x_{j-1})&=&\int_{-\infty}^{\infty } P_{\rm
Gauss}(s_{j}\sigma,x_{j}- x_{j-1}){\rm
exp}(-s_{j}^{2}){\frac{ds_{j}}{\sqrt{2\pi }}}
\label{dd}\\
P_{\rm Gauss}(\sigma,x)&=&{\frac{1}{\sqrt{2\pi } \sigma}}{\rm
exp}\left({-{\frac{x^{2}}{2\sigma ^{2}}}}\right)
\end{eqnarray}
The consideration of large and small values of \(s_j\) corresponds
to the possible presence or absence of eddies in that subregion at
various times. Note that the variance of both
\({\cal P}(x_{j}- x_{j-1})\) and \(P_{\rm Gauss}(\sigma,x)\) is
\(\sigma^2\).

(\ref{bb}) is constructed to give the cumulative effect of the
variable volatilities occurring in the various subregions. \(\rho\)
should be typical of scales on which viscosity influences and damps
the formation of eddies. As such
\(\rho\) is expected to lie between the dissipative (Kolmogorov)
scale, \(\eta\), and the lower end of the inertial scaling range (the
latter range being characterized by the absence of viscosity effects).
Note that if \(0<a<1\) then the size of the subregion,
\(r_{j+1}-r_{j}\), and thus the relevant eddy size for that subregion,
would increase with
\(j\).

We proceed by inserting (\ref{dd}) into (\ref{bb}) and integrating
over the \(x_j\). We can write the terms in the exponential which
depend on
\(y\equiv (x_{1},x_{2},...,x_{N})\) in matrix notation
\(y^{T}My+Jy\) where \(M\) is a matrix  and \(J\) is a vector, with
the latter depending on \(x_0\) and \(x_{N+1}\). We may complete
the square and do the Gaussian integrations over \(y\), for various
values of
\(N\). From this we are able to deduce a simple result,
where all dependence on the
\(s_j\) now resides in
\(\hat{S}^{2}\equiv \sum  _{j=1}^{N+1} s_{j}^{2}\). The \(s_{j}\)
integrations for fixed \(\hat{S}\) then just give the surface area of an
\(N+1\) dimensional sphere.
\begin{eqnarray}
\lefteqn{P_{N}(x_{N+1}-x_{0})}\nonumber\\
&=&{\frac{1}{\sigma (2\pi )^{{\frac{N}{2}}+1}}}\left[{\prod\limits_{j=1}^{N+1} \int_{-\infty }^{\infty }ds_{j}}\right]{\frac{1}{\hat{S}}}{\rm exp}\left({-{\frac{(x_{N+1}-x_{0})^{2}}{2\sigma
^{2}\hat{S}^{2}}}-{\frac{1}{2}}\hat{S}^{2}}\right)\nonumber\\
&=&{\frac{1}{\sigma (2\pi )^{{\frac{N}{2}}+1}}}\left({{\frac{2\pi
^{{\frac{N+1}{2}}}}{\Gamma
({\frac{N+1}{2}})}}}\right)\times
\int_{0}^{\infty }d\hat{S}\hat{S}^{N-1}{\rm
exp}\left({-{\frac{(x_{N+1}-x_{0})^{2}}{2\sigma
^{2}\hat{S}^{2}}}-{\frac{1}{2}}\hat{S}^{2}}\right)\nonumber\\
&=&{\frac{\left({(N+1)/2}\right)^{{\frac{N}{2}}}}{\sigma \sqrt{\pi
}\Gamma ({\frac{N+1}{2}})}}\int_{0}^{\infty }dSS^{N-1}{\rm
exp}\left({-{\frac{(x_{N+1}-x_{0})^{2}}{2\sigma
^{2}S^{2}(N+1)}}-{\frac{1}{2}}S^{2}(N+1)}\right)
\end{eqnarray}
In the last step we have defined \(S^{2}\equiv
\hat{S}^{2}/(N+1)\) for convenience.

Finally we will take this result and analytically continue from
integer values \(N+1\) to positive real values \(\tau\). Replacing
\(N+1\) by \(\tau\equiv (r/\rho)^a\) gives\footnote{This superposition of
Gaussians is reminiscent of the proposal in \cite{cd}.}
\begin{equation} P_{r}(\delta v_{r})={\frac{\left({\tau
/2}\right)^{{\frac{\tau -1}{2}}}}{\sigma \sqrt{\pi }\Gamma (\tau
/2)}}\int_{0}^{\infty}dSS^{\tau -2}{\rm exp}\left({-{\frac{(\delta
v_{r})^{2}}{2\sigma ^{2}S^{2}\tau }}-{\frac{1}{2}}S^{2}\tau
}\right).
\label{aa}\end{equation} We note that for large \(\tau\) the integral
over \(S\) becomes strongly peaked about \(S=1\), and thus
\begin{equation} P_r (\delta v_r)\stackrel{\tau \rightarrow \infty
}{\rightarrow} {\frac{1}{\sigma\sqrt{2\pi\tau }}}{\rm
exp}\left({-{\frac{(\delta v_{r})^{2}}{2\sigma^2\tau }}}\right).
\label{cc}\end{equation} In addition we find, {\em for any} \(\tau\),
that the variance in the velocity differences is
\begin{equation}
\left\langle{(\delta v_{r})^{2}}\right\rangle= \sigma^2\tau =\sigma
^{2}(r/\rho)^{a}.
\label{ee}\end{equation}

The variable \(\tau\) controls the evolution of the PDF, in the sense
that evolution through unit steps in \(\tau\) are generated by
successive application of the ``evolution operator'' in (\ref{dd}). In
this sense
\(\tau\) is an ``evolutionary'' time scale depending on the distance
scale
\(r\). There is also a dynamical time scale associated with scale \(r\),
the eddy turnover time, which is simply \(r/v\) where \(v\) is a typical
velocity observed on scale \(r\). If we associate \(v\) with
\(\sigma\sqrt{\tau }\) from (\ref{ee}), and if the
evolutionary and dynamical times happen to scale with \(r\) in the
same way, then we obtain \(a=2/3\). This gives the standard scaling
law \(\left\langle{(\delta v_r)^{2}}\right\rangle\propto r^{2/3}\).

We now perform the integral over
\(S\), and rescale the PDF such that the variance is unity for all
\(\tau\). We obtain
\begin{equation}
\hat{P}_{\tau }(x)={\frac{\tau ^{({\frac{1}{4}}(1+\tau
))}}{\sqrt{\pi }2^{{\frac{1}{2}}(\tau -1)}\Gamma
({\frac{1}{2}}\tau )}}{\left|{x}\right|}^{{\frac{1}{2}}(\tau
-1)}K_{{\frac{1}{2}}(\tau -1)}(\left|{x}\right|\tau ^{1/2}).
\label{ff}\end{equation}
\(K_\nu(y)\) is the modified Bessel function of the second kind. In
Fig. (1) we display
\(\hat{P}_{\tau }(x)\) for various \(\tau\).
\begin{figure}
\begin{center}
\includegraphics{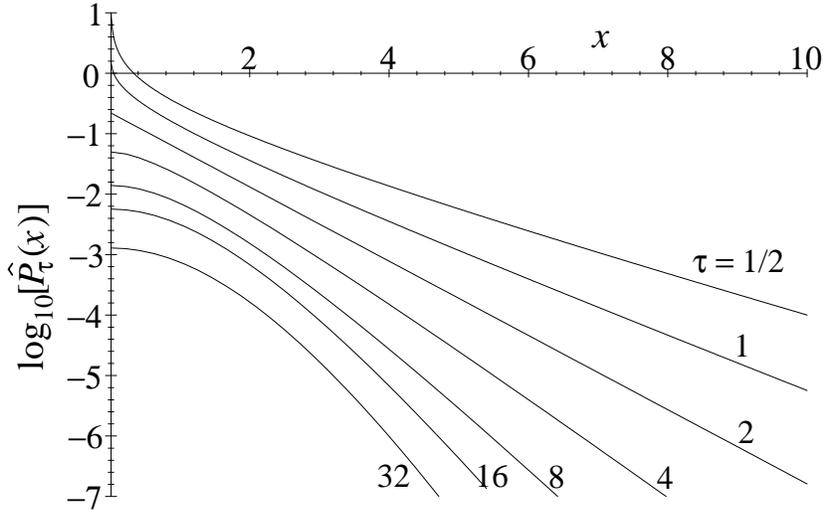}
\caption{\(\log_{10}[\hat{P}_{\tau }(x)]\) vs. \(x\) for
\(\tau=1/2,1,2,4,8,16,32\), top to bottom. The curves are vertically
displaced by \(0.5,0,-0.5,-1,-1.5,-2,-2.5\) respectively.}
\end{center}
\end{figure}

When \(\tau\) is an even integer, \(\hat{P}_{\tau }(x)\) is an
exponential times a polynomial in \(x\), and in particular for
\(\tau=2\) it is purely exponential, \(\hat{P}_2(x)={\rm
exp}(-\left|{x}\right|\sqrt{2})/\sqrt{2}\). For \(\tau<2\) the PDF
is even more strongly peaked and it decreases less quickly than an
exponential (``stretched exponential''), before becoming purely
exponential at large enough
\(x\). For  \(\tau>2\) we have ``deformed Gaussians'' with
exponential tails. For any \(\tau\),
\(d{\rm ln}\hat{P}_{\tau}(x)/dx\stackrel{x\rightarrow
\infty}{\rightarrow} -\sqrt{\tau}\). Alternatively we may consider
moments,
\(M_{\tau}^{n}\equiv
\int_{}^{}\hat{P}_{\tau}(x)\left|{x}\right|^{n}dx\), which are all
finite and given by
\begin{equation}
M_{\tau}^{n}={\frac{2^{n}}{\sqrt{\pi\tau^{n}}}}{\frac{\Gamma
({\frac{n+1}{2}})\Gamma ({\frac{n+\tau }{2}})}{\Gamma
({\frac{\tau }{2}})}}.
\end{equation} The approach to the exponential tails is reflected in
\(M_{\tau }^{n+1}/((n+1)M_{\tau}^{n})\stackrel{n\rightarrow
\infty}{\rightarrow}1/\sqrt{\tau}\).

It is common to describe the evolution of the PDF with
\(r\) in terms of the generalized structure functions, defined by
\(S_{n}(r)=\left\langle{\left|{\delta
v_{r}}\right|^{n}}\right\rangle\). For \(r\) in the inertial range it is
usually assumed that
\(S_n(r)\propto r^{\zeta_n}\). Our model suggests that exactly scale
independent exponents do not exist for the typical ranges of
\(r\) considered, and that instead we should consider the local
exponents
\(\zeta_n(r)\equiv d\log(S_n(r))/d\log(r)\). We obtain
\begin{equation}
\zeta_n(r)={\frac{a}{2}}\tau\left({\Psi({\frac{1}{2}}(n+\tau))-\Psi({\frac{1}{2}}\tau)}\right),
\label{ii}\end{equation} where \(\Psi\) is the digamma function.
This gives
\(\zeta_2(r)=a\), and in the large \(r\) limit, \(\zeta_n(\infty)=an/2\).
The latter is K41 scaling when \(a=2/3\). Although we adopt this
value of \(a\) in the following, we leave open the question of whether
\(a\) is exactly equal to \(2/3\) or just close to it.\footnote{For
example the first reference in
\cite{cc} advocates \(\zeta_2=.7\).}
We also stress that we are describing only the symmetric part of the
observed PDF, which in the case of the PDF for longitudinal velocity
differences also has an asymmetric part. This asymmetry is reflected
by the nonvanishing of the structure functions
\(\left\langle{\delta v_{\parallel r}^{n}}\right\rangle\)
for odd \(n\). In particular our
\(\zeta_3\) is not the same as the scaling exponent of
\(\left\langle{\delta v_{\parallel r}^{3}}\right\rangle\), which is
constrained to be unity \cite{cb,cz}.

In Fig. (2) we display
\(\zeta_n(r)\) for various \(r\), and compare to the large \(r\) limit
values
\(\zeta_n=n/3\). 
\begin{figure}
\begin{center}
\includegraphics{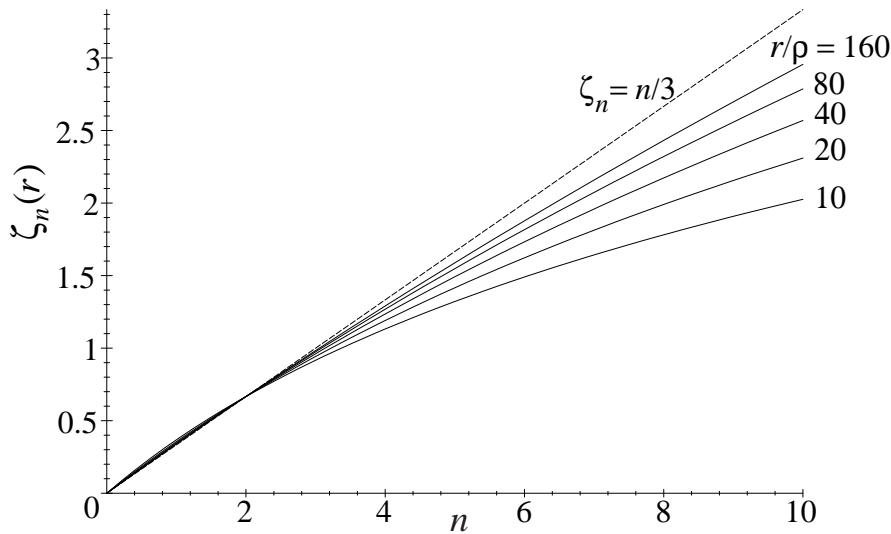}
\caption{\(\zeta_n(r)\) vs. \(n\) for \(r/\rho=10,20,40,80,160\),
bottom to top. The dashed line is \(\zeta_n=n/3\).}
\end{center}
\end{figure}
The main point is that the scaling exponents
approach their asymptotic values very slowly. It is also of interest that for some range of
\(r/\rho\), our exponents are realistic for a wide range of \(n\). We
show this by comparing to the model of She and Leveque
(SL)\cite{ce}, which yields \(\zeta
^{SL}_n=n/9+2-2\left({2/3}\right)^{n/3}\), and which is known to
fit the current data quite well. We may phrase the agreement in
terms of the relative scaling exponents
\(\zeta_n/\zeta_3\), which are known to better accuracy
\cite{cg,cf,ci,cc} than the individual exponents, and for which the
factor \(a\) in (\ref{ii}) cancels. We find,
for example, that
\(\zeta_n(35\rho)/\zeta_3(35\rho)\) as a function of \(n\) is within a
few percent of the corresponding SL values up to \(n=30\)! And as
\(r\) varies from 20 to 50, \(\zeta_6(r)/\zeta_3(r)\) for example varies
from 1.72 to 1.82, compared to
\(\zeta^{SL}_6/\zeta^{SL}_3=1.78\).

To see the \(r\) dependence of the structure functions themselves we
display \(S_n(r)\) in Fig. (3).
\begin{figure}
\begin{center}
\includegraphics{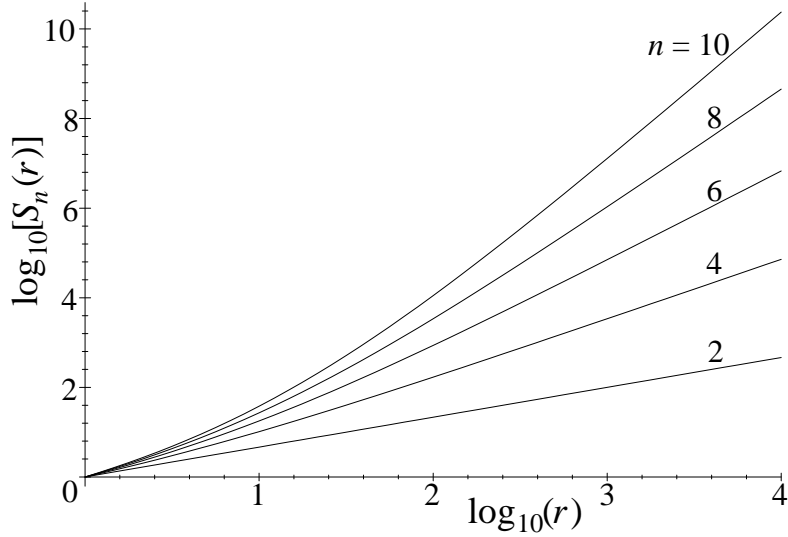}
\caption{\(\log_{10}[S_n(r)]\) vs. \(\log_{10}(r)\) with \(\rho=1\) for
\(n=2,4,6,8,10\), bottom to top, each line vertically offset so as to
vanish at \(r=1\).}
\end{center}
\end{figure}
Over some range of
\(r\) the lines appear to be close to straight. But other than for
\(n=2\), which is exactly straight, the lines have a positive curvature
which increases for larger
\(n\). We will see below how these
positive curvatures, or equivalently the scale dependence of the
scaling exponents, may be rather subtle to observe experimentally.
But first we must consider the effects of viscosity on the structure
functions.

We expect that the effects of viscosity will cause \(\tau=(r/\rho
)^{2/3}\) to be replaced by a function \(f(r)\) which deviates from
\((r/\rho )^{2/3}\) as \(r\) approaches the dissipative scale
\(\eta\) from above. Since a decreasing \(\tau\) corresponds to
increasing intermittency, and since the cause of intermittency in our
picture---variable volatility of velocity differences---should be
damped by viscous effects, we expect that \(f(r)\) should start to
decrease more slowly than \((r/\rho )^{2/3}\). Thus the evolution of
the PDF shapes is retarded by viscous effects in the ``intermediate
viscous range'', the range of \(r\) between
\(\eta\) and the start of the inertial range. The implication is that at
large scales in the inertial range where
\(f(r)=(r/\rho )^{2/3}\) it is likely that \(\rho>\eta\). We may also
expect that \(\rho/\eta\) grows with the size of the intermediate
viscous range.

For very small \(r\) where the velocity field is smooth the variance is
evolving like \(r^2\), faster than in the inertial range, and thus there
must clearly be a decoupling between the evolution of the PDF
shapes and the evolution of the PDF variance. We are thus led to a
generalization of our model, where we replace the expression in
(\ref{aa}) by the following.\footnote{This expression could be ``derived'' from a path
integral as before, where the discretization of the range from 0 to
\(r\) would be determined by unit steps in the function \(g(r)\).
Instead of each point, groups of adjacent points would have a single
fluctuating volatility, and the transition from one grouping to the
next would determined by unit steps in the function \(f(r)\).}
\begin{equation} P_{r}(\delta
v_{r})={\frac{\left({f(r)/2}\right)^{(f(r)-1)/2}}{\sqrt{\pi g(r)/f(r)
}\Gamma (f(r)/2)}}\int_{0}^{\infty }dSS^{f(r)-2}{\rm
exp}\left({-{\frac{(\delta
v_{r})^{2}}{2S^{2}g(r)}}-{\frac{1}{2}}S^{2}f(r)}\right)
\label{gg}\end{equation} This new PDF has the property that
\(\left\langle{(\delta v_{r})^{2}}\right\rangle= g(r)\), and thus we
can use \(g(r)\) to reproduce the observed behavior of the variance
even in regions where viscosity or finite size effects are important.
The function
\(f(r)\) determines the evolution of the PDF shapes; that is, the new
\(\hat{P}_{r}(x)\) is obtained from the one in (\ref{ff}) by replacing
\(\tau\) with \(f(r)\).

We will illustrate the effects of viscosity with specific choices for
\(f(r)\) and \(g(r)\). We consider \(f(r)=\left({[r+\rho -\eta]
/\rho}\right)^{2/3}\), which is unity at \(r=\eta\) and approaches
\((r/\rho)^{2/3}\) at large \(r\).\footnote{Although we expect \(f(\eta)\)
to have some dependence on the size of the inertial range, we take
\(f(\eta)=1\) for simplicity.}  For
\(g(r)\) in the range \(\eta<r<10^4 \eta\) we consider
\[g(r)=\left[{c_1-c_2e^{[1-{\frac{r}{\eta
}}]/c_4}-c_3e^{[1-10^{4}{\frac{\eta
}{r}}]/c_5}}\right](r/\eta)^{2/3}.\]
We take two examples for
\((c_1,c_2,c_3,c_4,c_5)\) which idealize typical data sets,
\(g_1(r)\) with \((80,79,60,10,5)\) having a relatively large inertial
range (large Reynolds number) and
\(g_2(r)\) with \((80,79,70,5,30)\) having a small inertial range. Our
results are not very sensitive to precisely how these functions
deviate from
\(r^{2/3}\) behavior outside the inertial range. Note that the
intermediate viscous range is larger for \(g_1(r)\).\footnote{The
tendency for the intermediate viscous range to grow with the
Reynolds number has been noted for example in the first reference
in \cite{ci}.} Due to
our expectation that
\(\rho\) should increase with the size of the intermediate viscous
range it is natural that
\(\rho_1>\rho_2\). For illustrative purposes only, we choose
\(\rho_1=4.3\eta\) and \(\rho_2=1.3\eta\).

With this input we can extract all higher order structure functions
from (\ref{gg}). The scaling in the inertial range for both cases turns
out (because of our fortuitous choice of
\(\rho_1\) and \(\rho_2\)) to be described by the realistic scaling
exponents
\(\zeta_n(35\rho)\) from (\ref{ii}). We show this in Fig. (4)
where we plot
\(S_n(r)/r^{\zeta_n(35\rho)}\).\footnote{This figure may be
compared with Fig. (1) in the second reference of \cite{cc}.}
\begin{figure}
\begin{center}
\includegraphics{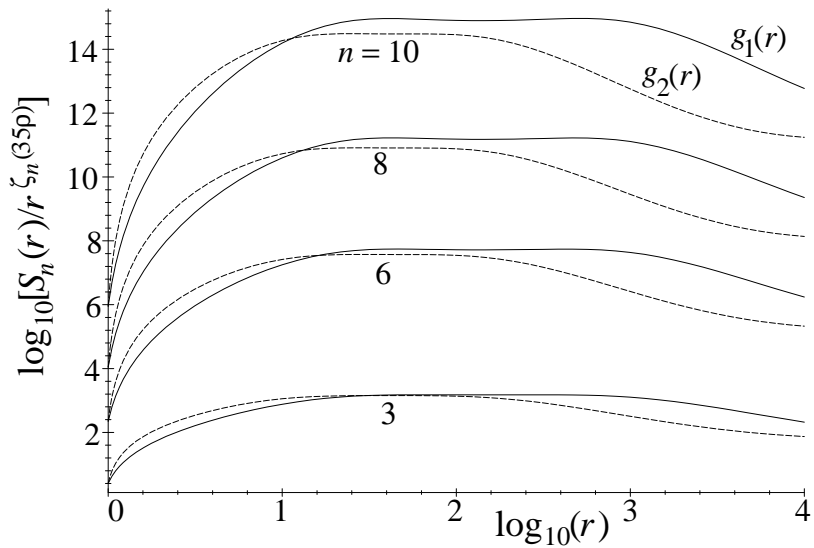}
\caption{\(\log_{10}[S_n(r)/r^{\zeta_n(35\rho)}]\) vs.
\(\log_{10}(r)\) with \(\eta=1\) for \(n=3,6,8,10\), bottom to top. The
solid and dashed lines correspond to \(g_1(r)\) and \(g_2(r)\)
respectively.}
\end{center}
\end{figure}
Of most
interest are the
\(n=8,10\) curves for the
\(g_1(r)\) case, where we see (very slight!) evidence of a double
hump structure growing more prominent with increasing \(n\). This
positive curvature in the middle of the scaling region is arising from
the positive curvatures in Fig. (3), which in turn is a
reflection of the gradual approach to K41 scaling. This is the generic
signature of our model, which should be seen when the size of the
inertial range and the order
\(n\) are both large enough. The figure makes it clear, though, that it
is a small effect which may be hidden in present data.

Our model may also be used to illustrate ``extended self-similarity''
\cite{ck,cc}. In Fig. (5) we plot \(S_n(r)\) versus
\(S_3(r)\) for \(4<r/\eta<10^4\) for the two cases \(g_1(r)\) and
\(g_2(r)\).
\begin{figure}
\begin{center}
\includegraphics{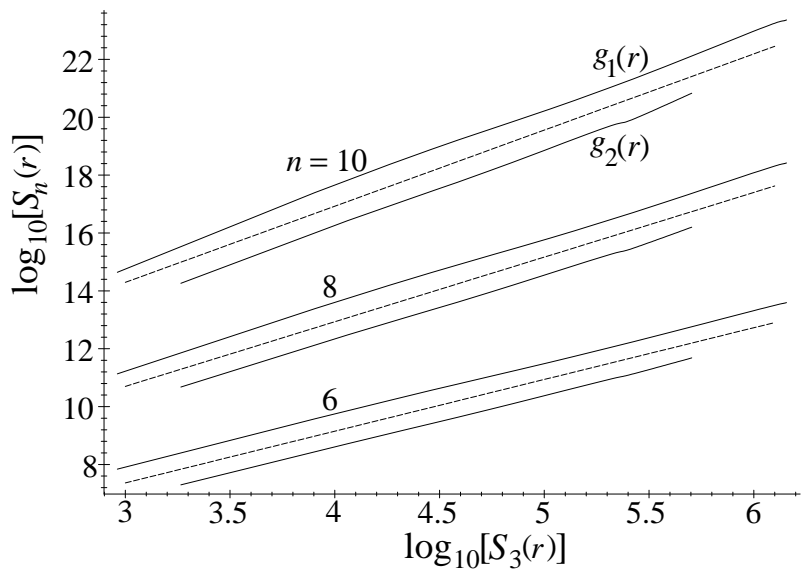}
\caption{\(\log_{10}[S_n(r)]\) vs.
\(\log_{10}[S_3(r)]\) for \(4<r/\eta<10^4\)
and for \(n=6,8,10\), bottom to top. The lines for the \(g_1(r)\) and
\(g_2(r)\) cases are displaced by 0.5 and \(-0.5\) respectively.
The dashed straight lines have slopes
\(\zeta_n(35\rho)/\zeta_3(35\rho)\).}
\end{center}
\end{figure}
For comparison we add straight lines with slopes given by
the relative exponents
\(\zeta_n(35\rho)/\zeta_3(35\rho)\). We see that scaling has been
extended to smaller scales than is apparent in Fig. (4). Similar
results are obtained for other choices of the functions
\(f(r)\) and \(g(r)\). It thus appears that such plots are not very
sensitive to the deviations from power law scaling we are proposing.

We reiterate that the model yields a
universal set of {\em local} scaling exponents, and the scaling
exponent from an experiment depends on what distance scale,
effectively, the local scaling exponent is being measured at relative
to \(\rho\). On the other hand if it is true that
\(\rho/\eta\) increases with the size of the intermediate viscous
region, as we are suggesting, then it may be difficult to obtain
measurements at distances, in units of \(\rho\), which are very
different from each other. The variability in the scaling exponents is
also obscured if different experiments have intermediate viscous
regions of similar size, and/or if one is confined to the lower order
structure functions, such as \(n=6\) and below.

We have seen how viscosity and finite size effects can
have the effect of transforming the structure functions in Fig.
(3) into the structure functions in Fig. (4) which in turn
display extended self-similarity in Fig. (5). It is also
encouraging to find that realistic values of the scaling exponents
emerge when the
\(\rho\) parameter is within the intermediate viscous range.
But perhaps most important is that the model suggests how evidence
for a slow approach to K41 scaling could eventually be uncovered.
We should also differentiate between the evolution of the PDF
shapes as a function of
\(r\) (to which the structure functions are sensitive), and the basic set
of shapes predicted by the model, given in (\ref{ff}) and depicted in
Fig. (1).  These probability density functions may be of
interest in various other contexts.

\section*{Acknowledgments} I thank Brian Smith for
discussions. This research was supported in part by the Natural
Sciences and Engineering Research Council of Canada.

\end{document}